\documentclass[prl,twocolumn,english,aps]{revtex4}
\usepackage[T1]{fontenc}
\usepackage[latin9]{inputenc}
\usepackage{amsmath}
\usepackage{amssymb}
\usepackage{graphicx}
\usepackage{esint}

\makeatletter

\@ifundefined{textcolor}{}
{%
 \definecolor{BLACK}{gray}{0}
 \definecolor{WHITE}{gray}{1}
 \definecolor{RED}{rgb}{1,0,0}
 \definecolor{GREEN}{rgb}{0,1,0}
 \definecolor{BLUE}{rgb}{0,0,1}
 \definecolor{CYAN}{cmyk}{1,0,0,0}
 \definecolor{MAGENTA}{cmyk}{0,1,0,0}
 \definecolor{YELLOW}{cmyk}{0,0,1,0}
}

\usepackage[normalem]{ulem}\@ifundefined{definecolor}{\usepackage{color}}{}

\usepackage[version=3]{mhchem}
\usepackage{babel}

\makeatother

\usepackage{babel}% Utilizamos el paquete para gestionar imagenes jpg
\usepackage{graphicx}

\usepackage{amsfonts}
\usepackage{textcomp}
\usepackage{float}
\usepackage{amsmath}
\usepackage{cases}
\usepackage{mathtools}

\graphicspath{%
    {converted_graphics/}% inserted by PCTeX
    {/}% inserted by PCTeX
    {C:/Users/Emilio/Documents/MATLAB/TESIS MAESTRIA matlab/unidimensional/flujo nulo/}% inserted by PCTeX
    {C:/Users/Emilio/Documents/MATLAB/calciowave/}% inserted by PCTeX
}
\begin{document}
% Definimos titulo, autor, fecha.
\title{Multiscale modeling of exocytosis in the fertilization process}
%{Multiscale modeling of early events of fertilization in  \emph{urchin sea} eggs}
\author{Aldo Ledesma Durán, I. Santamaría-Holek}
\affiliation{UMDI-Facultad de Ciencias, Universidad Nacional Autónoma
de México Campus Juriquilla, 76230, Querétaro, Mexico}

\vspace{7mm}
{\color{blue}\date{Actualización: 8 julio 2014} }    % Enter 
% Generamos titulo e indice de contenidos

%%%%%%%%%%%% nuevo archivo
\begin{abstract}
We discuss the implementation of a multiscale biophysico-chemical model able to cope 
with the main mechanisms underlying cumulative exocytosis in cells. The model is based
on a diffusion equation in the presence of external forces that links calcium signaling and the 
biochemistry associated to the activity of cytoskeletal-based protein motors. This multiscale model offers an 
excellent quantitative spatio-temporal description of the cumulative exocytosis measured by means
of fluorescence experiments. We also review pre-existing models reported in the literature on calcium 
waves, protein motor activation and dynamics, and intracellular directed transport of vesicles.
As an example of the proposed model, we analyze the formation of the shield against polyspermy in 
the early events of fertilization in  \emph{sea urchin} eggs.

\vspace{3mm}
\end{abstract}

\maketitle

\section{Introduction}
%\section{Multiscale modeling}

Biological processes frequently involve several mechanisms and sub-processes that link different temporal and length scales such as, for example, intracellular molecular interactions to the scale of cells and beyond to the behavior of collectives of cells and even organisms \cite{Alberts}. From an applied point of view, 
biomedical research often performs experiments in which these scales are also imbricated \cite{Saltzman}. 
The quantitative modeling of these systems and the processes they perform often requires the extensive use of computational 
simulations that have to incorporate several simulation techniques, each one according to the time and
space scale associated to the biological process in question \cite{NIH}. 

However, simple theoretical 
models able to cope with these problems are less developed in the literature, although they may be of
great importance because they can simplify the quantitative description without loss of accuracy and,
more important, by allowing to give clear interpretations of the different mechanisms in terms of simple and
well understood physicochemical laws.

In this short review, we aim to describe, link and use three theoretical models reported in the literature that
allow to formulate a theoretical multiscale model able to cope, with high precision, with intracellular
processes involving exocytosis, that is,  with the process by which a cell directs 
the contents of secretory vesicles out of the cell membrane and into the extracellular space. We 
look at this particular biological process due to its ubiquity and biomedical importance ranging from transcelular transport,
insulin, neurotransmitter and enzyme secretions in the energetic, nervous and early developmental
metabolic processes and pathways \cite{keener1998}. In particular, we will study the case of the early events of
fertilization, that is, the formation of the shield against polyspermy in oocytes, see Figure \ref{vitaline}.

%%%%%TEXTO AÑADIDO

Exocytosis is a very broad and complex process that involves different sub-processes such as the formation and storage of vesicles in different pools inside the cell, the calcium signaling process, the active transport, and other related processes such as vesicle docking, priming and the final fusion of the vesicle in the porosomes of the plasmatic membrane\cite{anderson, jena}. Many of these processes have been studied experimentally and, as a consequence, several proteins and enzymes have been discovered. However, beyond the knowledge of their existence and participation, there is still a fundamental lack of understanding concerning the specific role they play in the secretion process. \cite{jahn}.

Depending on the particular physiological function of the exocytosis in each cell type, there are variations in the rate of exocytosis, the lag time before it begins, its time course, the proportion of vesicles that undergo fusion in response to stimulation, the presence of single or multiple granule types, and the nature of the regulation of exocytosis by second messenger pathways \cite{burgoyne}.  For example, in the particular case of fertilization of urchin sea eggs, the exocytotic process involves sperm fusion with cell membrane, the influx of external calcium and the subsequent elevation of calcium concentration due to the internal pools. Subsequently, other fundamental processes take place like the formation and storage of cortical granules, the docking, priming and fusion of the secretory vesicles with the plasma membrane, presumably through SNARE proteins. Finally, the formation of the vitaline envelope and the final endocytosis and recycling of empty vesicles are processes that may play an important role in regulating the exocytosis \cite{wong}.

The presence of micotubules and actin filaments may play opposite roles depending on the cell type. In some cases, the excytosis becomes favored by the presence of the cytoskeleton, as in the case of serotonin and hyaline secretion \cite{biophysics}, whereas in other cases it may be inhibited by the presence of the cytoskeleton that acts as a mechanical barrier \cite{oheim, lang}. Presumably, cytoskeleton based secretion depends on the action of protein motors whereas the opposite case suggest a diffusion controlled exocytosis \cite{santamaria2009}. The distances travelled by the secretory vesicles in both cases are very different. In this review, we focus on the first case, that is, in which secretion is mediated by cytoskeleton associated protein motors. Therefore, for the particular case of the urchin sea fertilization, we have considered calcium signaling, vesicle diffusion and protein motor activation and its processivity as the rate controlling steps in the secretion process.

\begin{figure}[]
\centering
\includegraphics[scale=0.3]{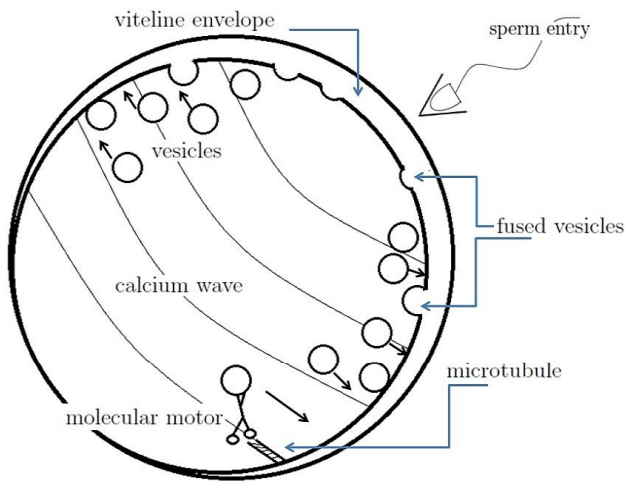}
\caption{Scheme showing the three main mechanisms associated to the initial events of oocyte fertilization. 
Sperm entry through the acrosomic reaction liberates the genetic material into the oocyte and triggers the local 
entrance of external calcium and the subsequent propagation of a calcium wave through the ooplasm. This 
process activates protein motors, that actively transport clusters of cortical vesicles towards the plasmatic 
membrane. Once fused, hyaline is released and initiates the gradual activation of the vitaline envelope until 
the shield against polyspermy is completed.}
\end{figure}\label{vitaline}

An appropriate measure of the cumulative exocytosis 
by release site can be performed with the help of fluorescence experiments \cite{gqbi1997, baker1978,biophysics}.
In these experiments, a fluorescent dye is dissolved in the culture medium in such a way that it is activated 
when becomes adsorbed in the plasma membrane and is illuminated with the appropriate light. During secretion,
intracellular vesicles fuse with the plasmatic membrane to release its contents. Exposed to the culture medium, the
membrane of the vesicles adsorbs the dye and increases the local fluorescence, see Figure \ref{cumulative}.
\begin{figure}[]
\centering
\includegraphics[scale=0.5]{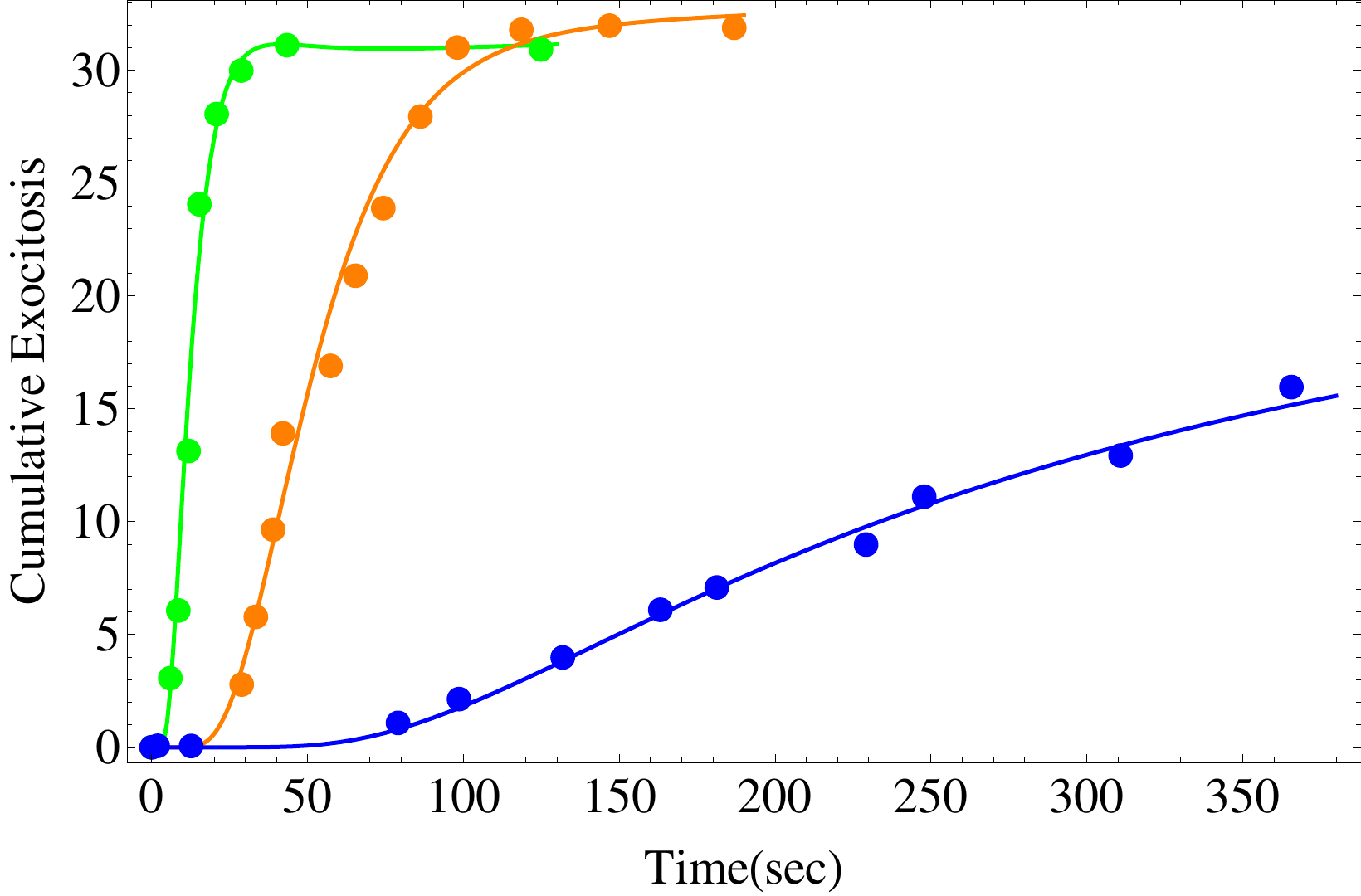} 
\caption{Sample traces of exocytosis in eggs for different external calcium concentrations. For each series, the
number of newly emerged exocytosis disks in each image frame
was counted. This number plus the numbers from all previous
frames gave the cumulative number for this time point. Each line correspond to different values of external calcium: 0.4, 2 and 10mM for blue, orange and green points respectively. For fitting we use $\beta=1\times 10^{-6} s^{-1}$, $w=223.6 s^{-1}$ in equation (\ref{eqs:fp}). Other parameters appear in the box inside the text.}
\end{figure}\label{cumulative}

In view of this, the theoretical description of cumulative exocytosis is based on an apparently simple model.
The main question is to know the number of vesicles that arrive to a given release site of area $A$ in 
the membrane and fuse with it as a function of time.  Therefore, if $\mathbf{J}(t)$ counts the number 
of cortical granules per unit area that reach the release site at the membrane per unit time, 
%(of radius $\nu$), 
then the number $n(t)$ of fused cortical granules as a function of time is given by the relation
\begin{equation}\label{Eq:1}
n(t)=  A \xi \int_0^t \mathbf{J}(d,t')dt',
\end{equation}
where $A$ is the measure area and $d$ the initial position of the vesicle cluster with respect to the 
plasmatic membrane \cite{baker1978}. The fusion efficiency parameter $\xi$ is the fraction of vesicles that release its content with respect to the total number of vesicles arriving at the plasma membrane. Here, for simplicity, we will assume that this parameter is the unity.
In principle, this simple model quantifies the cumulative exocytosis 
observed in experiments \cite{gqbi1997, baker1978}. 

However, it is clear that the current $\mathbf{J}(t)$ must be determined explicitly in order to make a 
quantitative progress. Determining $\mathbf{J}(t)$ is not a trivial task.
The mechanism of the secretion of the vesicular contents out of the cell is a complex process that involves
calcium signaling, activation and action of several cytoskeletal motors through distances typically involving
few microns. This is an energy consumption process due mainly to the activity of the protein
motors and pumps. Therefore, this makes very important to consider the detailed description of the dynamics and 
energetics of these entities.

In fact, in order to obtain $\mathbf{J}(t)$ and  perform the program outlined, one needs to formulate a 
multiscale model that accounts for the initial events of 
fertilization. In particular, we have to reproduce the reaction-diffusion wave 
propagation of internal and external calcium and couple it with a directed diffusion model describing the 
transport of granule clusters containing hyaline to the plasma membrane. This second step considers 
in parallel the biochemical kinetics of protein 
motors, kinesins, myosins and dyneins \cite{Sperry}. 
As a consequence, our work gives quantitative evidence that the long time enzyme secretion of cortical vesicles 
or granules is driven by protein motors, mostly by kinesins.

The paper is organized as follows. In Section 2 we review previous theoretical works on reaction-diffusion 
calcium signaling; protein motor biochemistry, energetics and dynamics; and finally on diffusion 
in the presence of external forces, that allow to propose an unifying view of the exocytotic processes
from the a physical chemistry point of view. In Section 3  we apply the multiscale model 
to describe the slow formation of the shield against polyspermy, which occurred when exocyted vesicles 
lift the viteline envelope away. In particular, we join the results of the pre-existing models on calcium 
waves, protein motors activation and intracellular directed transport of vesicles in a cell for 
a successful quantification of cumulative exocytosis previously reported in experimental data. 
Finally, Section 4 is devoted to present our main conclusions.

\section{Review of theoretical backgrounds}
As we have already mentioned, the exocytosis of substances by cells comprise several mechanisms
whose interplay is essential. This interplay links several time and length scales that should be modeled
in a nested way. Looking at the events, the entrance of external calcium
into the cell through the corresponding ion channels and its interaction with the intracellular calcium
is the triggering one of subsequent processes, like
activation of cytoskeletal-based protein motors and their active transport of vesicles and granules
towards the plasmatic membrane along microtubules. Here, we review previous theoretical work reported on 
these three main processes.

In order to fix ideas, we will discuss the general aspects of the three models by considering the phenomenology 
associated to the fertilization of sea urchins eggs. This may help the reader to create a clearer image of 
the secretion process and the mechanisms involved.

%%%%%%%%%%
\subsection{Locally-constrained and directed diffusion}\label{sec:difusion}
%%%%%%%%%%

The mechanisms participating in exocytotic processes that we already discussed, 
calcium wave signaling and protein motor action converting chemical energy into mechanical one, 
can be linked by means of a diffusion model in the presence of external forces. Because
this diffusion equation is very general, we will first discuss it in order to show where the
other mechanisms enter.

Mass transport inside cells is always affected by thermal agitation
that induces a diffusive motion. This diffusion is more or less important depending on the external
forces acting over the transported bio-particles. In addition, the intracellular medium consists of a wide 
variety of polymers and organelles whose presence causes  a viscoelastic behavior of the cytosol which,
in principle, should be taken into account \cite{gittes}. As a consequence of this, when
bio-particles undergo different forms of passive diffusion in the intracellular medium they often
show anomalous diffusion  (subdiffusion)  \cite{weitz,JCPfinite}. 

Subdiffusion has been observed in the passive transport of particles in the cellular medium whereas an enhanced diffusion has been reported for particles driven by protein motors  \cite{caspi2000}. 
This enhanced diffusion was analyzed in terms of a generalized diffusion equation containing forces due to the cytoskeleton network and to the protein motors \cite{biophysics,santamaria2009,santamaria2014}.
However, when protein motors carry bio-particles of different sizes, this anomalous diffusion is sometimes
weak and normal diffusion can be used to describe the motion with an effective diffusion coefficient. 

Local confinement and protein motor action are therefore the key ingredients to explain vesicle transport in 
cumulative exocytosis \cite{biophysics,santamaria2014}.

Let us consider $\rho(\mathbf{x},t)$ as the number of transported vesicles per unit volume 
and $D_v$ is the diffusion coefficient of vesicles in the cytoplasm. Then, the time evolution 
of $\rho(\mathbf{x},t)$  is described by the general equation
\begin{equation}\label{eqs:fp}
\frac{\partial \rho}{\partial t}(\mathbf{x},t)= D_v \nabla^2 \rho 
- \frac{1}{\beta} \nabla \cdot \left[  \rho \mathbf{F} \right]
\end{equation}
where $\mathbf{F}$ is a vector which represents the total force over the transported 
vesicles and $\beta$ is the friction coefficient of the transported vesicles.  
Here, it has been proposed that force $F$ could be a combination of harmonic radial 
forces \cite{biophysics,santamaria2014,santamaria2009}: 1) a locally constraining force that keeps the vesicles fixed to 
microtubule an 2) a pulling force due to the activity of molecular motors. 

Thus, the first force can be modeled as resulting from a local harmonic force $\mathbf{F}=- w^2 \mathbf{x}$ 
which is applied simultaneously with the force of the motors that move with velocity $v$ \cite{aldo2008,santamaria2014,biophysics}.
The total force applied on the vesicles is an harmonic force displacing in the same direction as the molecular
motors with velocity $v(t)$. As a consequence of this, it may be modeled by means of the relation
\begin{equation}\label{eqs:force0}
F= - w^2\left[\mathbf{x}-\mathbf{x}_{motor}(t)\right]
\end{equation}
where $\mathbf{x}_{motor}(t)$ represents the position of the motor as a function of time \cite{aldo2008,biophysics}. 
Hence, if the protein motor velocity is, in general, time dependent, then the force takes the form
\begin{equation}\label{eqs:force}
\mathbf{F}(\mathbf{x},t)=-w^2\left[(x-x_o)- \int_{t_0}^t  v(t')dt'\right]\mathbf{e}_r,
\end{equation}
where $x_o$ is the initial position of the transported vesicles with respect to the release site.
In writing Eq. \ref{eqs:force}, we have supposed that:  1) mobilization of motor does not occurred until some critical time  $t_0$ when local Ca$^{2+}$ has reached some critical value; 2) before its activation, kinesin exerts a trapping force $-w^2(x-x_0)$ over the vesicles, and 3) once activated, the movement and influence of the kinesin is only radially outward, being $\mathbf{e}_r$ the outward unity vector.
Calcium waves and the biochemistry of molecular motors, that will be discussed in the following subsections,
allow to determine an explicit expression for the average motor velocity $v(t)$ and its activation time.
In general, we may assume that the process starts instantaneously when there exist saturation levels 
of ATP and local Ca$^{2+}$ concentration increases over a threshold value. This situation 
can be taken into account through a Heaviside function in the form $v(t) \rightarrow \Theta(t-t_i)v(t)$,
where $t_i$ depends on time and position through Ca$^{2+}$ concentration.

Inserting Eq. (\ref{eqs:force}) in (\ref{eqs:fp}), one can find $\rho$ using the appropriate initial and boundary conditions. 
The experiments suggest that one may use three distinct pools of vesicles as initial condition \cite{gqbi1997}: one very near membrane, other in the center of the cell and one in intermediate position. As a boundary condition, one may  
suppose that vesicles reaching the membrane fuse to the plasmatic membrane at the same rate at which they arrive.

From the previous equations, we can know the number of cortical granules that reach cell membrane during exocytosis. 
To obtain this quantity, one must calculate the current of vesicles $\mathbf{J}(\mathbf{x},t)$ implicitly defined by the conservation equation $\rho_t=\nabla\cdot \mathbf{J}$.  Direct comparison with equation (\ref{eqs:fp}) reveals that the 
current is given by
\begin{equation}
\mathbf{J}(\mathbf{x},t)= - D_v \nabla \rho + \beta^{-1}\rho\mathbf{F}.
\end{equation}

Since we are interested in cumulative exocytosis, we need to sum each vesicle arriving to the membrane as a 
function of time. If $\mathbf{J}(d,t)$ counts the density of cortical granules reaching the membrane from the initial
position $d$ of the cluster, then the number $n(t)$ of fused vesicles is given by the relation (\ref{Eq:1}).

As we have mentioned above, we need now to determine the time at which the motors are activated and the
time dependence of the molecular motors. In order to see how this can be done, in the following subsections
we will discuss the role and phenomenology of calcium waves and signaling, and of the biochemistry and dynamics
of the molecular motors.

%%%%%%%%%%
\subsection{Calcium waves}
%%%%%%%%%%
Two main models  are  widely used to explain intracellular oscillations
and waves: 1) the calcium induced calcium release model (CIRC) and 2) the two-pool model (TPM). 
Both models are based on reaction-diffusion equations for the Ca$^{2+}$ concentration and one for 
the calcium stored in the internal reservoirs Ca$^{2+}_i$. The election of the model should be based in accordance with
cell type and the kind of stimulation. By solving these models with the appropriate boundary condition, 
one obtains the spatio-temporal behavior of the Ca$^{2+}_i$ concentration. 

Inside the cells, after the external stimulation, oscillations of the calcium concentration produce 
spatial non-homogeneities along the cell. These non-homogeneities seem to induce a spatio-temporal 
organization of cell's activity, that is, calcium waves act as a signal starting different processes at
different times and positions, and is a function of the initial stimulation. 

In vertebrates, the Ca$^{2+}$ is mainly stored in bones, from where it may be released to the blood
after the adequate hormonal stimulation. Inside many types of cells, Ca$^{2+}$ it is stored mainly 
in the endoplasmatic reticulum. This allows to use it when necessary. Intracellular Ca$^{2+}$ concentration 
is lower than the extracellular one, $0.1\mu M$ and $1 \mu M$, respectively. This implies that the 
entrance of Ca$^{2+}$ through the corresponding ion-channels after membrane depolarization is an 
spontaneous irreversible process. This means that no energy is necessary to add in order to perform this entrance.
After the response of the cell to the stimulation is completed, molecular pumps regulate again the Ca$^{2+}$
concentration inside the cell. This one is an energy consuming process which is necessary because Ca$^{2+}$ may be toxic for high concentrations in blood and cells.
 
Specifically, the Ca$^{2+}$ is removed from the cytoplasm 
in two ways: 1) pumping it out of the cell and 2) pumping it inside several organelles 
like the mitochondria, the endoplasmic reticulum (ER) or
the sarcoplasmic reticulum (SR). Thus, during the initiation and the propagation of 
a calcium wave, the influx of Ca$^{2+}$ occurs in two ways: 1) the entrance of extracellular calcium and 2) the release of calcium 
from the internal reservoirs. 

The two internal calcium reservoirs are activated by two types of receptors:
ryanodine and IP$_3$.  With the attachment of these receptors on the membrane,  the ion-channels open allowing the release of calcium to the cytosol. Ryanodine is more common in neurons and pituitary cells
whereas IP$_3$ is mainly found in non-muscular cells.

%%%%%%%%
\paragraph{Calcium-Induced Calcium-Release model.}
%%%%%%%%
The calcium-induced calcium release-model (CIRC) was originally proposed
by Friel in Ref. \cite{friel} and it is more frequently use in
ryanodine dependent processes.  
\begin{figure}[] % float placement: (h)ere, page (t)op, page (b)ottom, other (p)abe
  \centering
  \includegraphics[width=3.25in,height=2.21in,keepaspectratio]{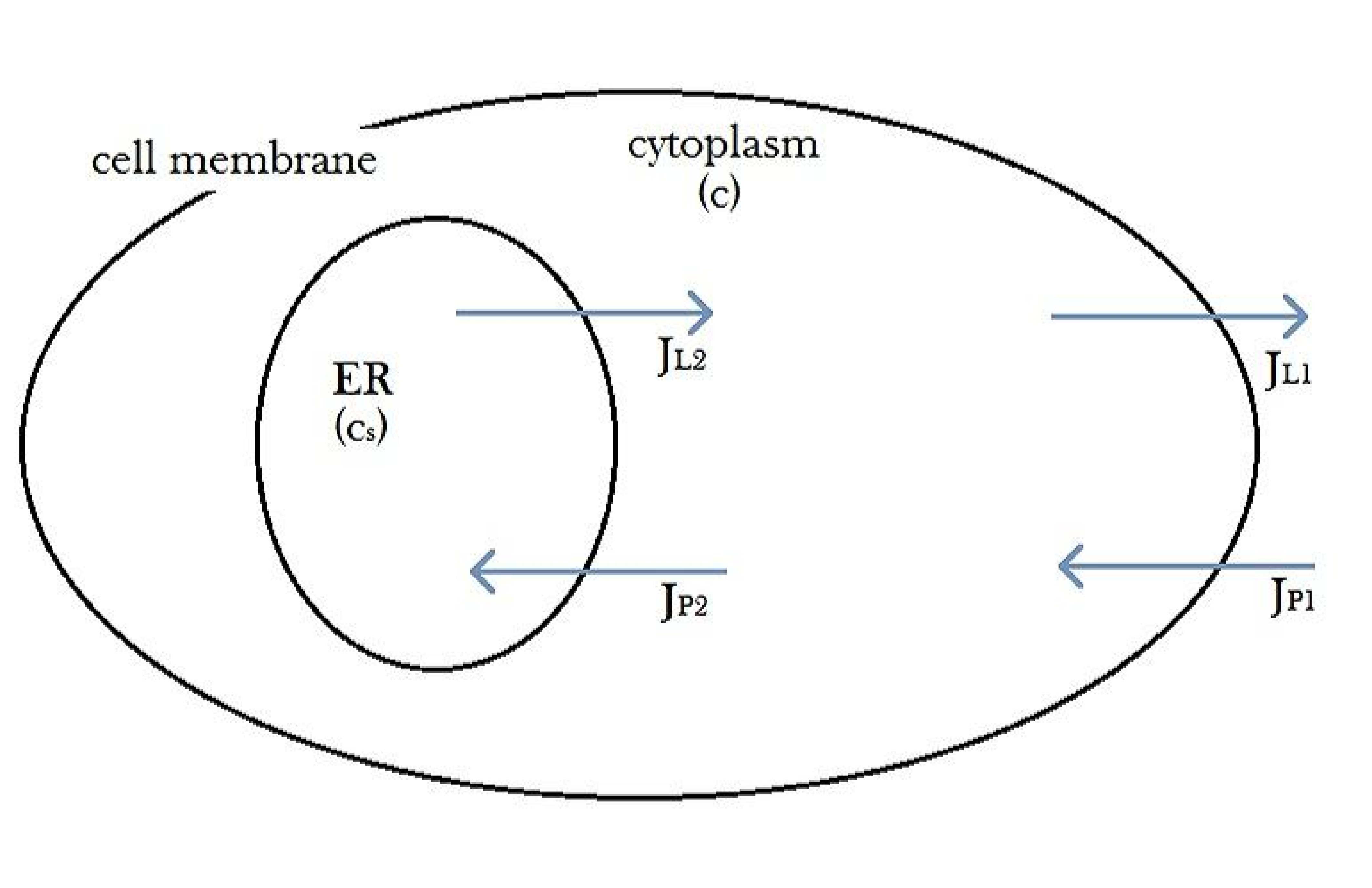}
  \caption{Flux diagram for the CICR model. In this case, a single calcium reservoir
  is activated due to the increase of internal calcium concentration.}
  \label{fig:cicr}
\end{figure}

The CIRC model is based on a balance of Ca$^{2+}$ fluxes in the cell
as it is shown in Figure 2. A single internal reservoir (ER) exchanges Ca$^{2+}$
with the cytosol through the fluxes $J_{L2}$ and $J_{P2}$. The cytosol, in turn
exchanges calcium with the extracellular environment with the fluxes  $J_{L1}$ y $J_{P1}$.
Denoting the calcium Ca$^{2+}$ concentration in the cytoplasm by $c$ and $c_s$ the 
calcium concentration in the internal reservoir Ca$^{2+}_i$, the model postulates the following
dynamics

\begin{eqnarray}\label{calcio1}
 \frac{d c}{dt} &=& J_{L1}-J_{P1}+J_{L2}- J_{P2},   \\
\frac{d c_s}{dt}&=& -J_{L2}+ J_{P2},
\end{eqnarray}

where the fluxes depend on the concentrations in the following way:
%\[ \begin{array}{lcl}
\begin{eqnarray}\label{calcio2}
J_{L1}=&k_1 (c_e-c), &\hspace{1cm}\mbox{Entrance}. \nonumber \\
J_{P1}=&k_2 c,      &\hspace{1cm}\mbox{Exclusion}. \nonumber\\
J_{L2}=&k_3 (c_s-c),&\hspace{1cm} \mbox{Release}. \\
J_{P2}=&k_4 c,      &\hspace{1cm}\mbox{Absortion}. \nonumber 
\end{eqnarray}
%\end{array}\]
Here, $c_e$ is the external  Ca$^{2+}$ concentration that may be assumed as constant due to
the large concentration difference. However, because a linear model do not leads to the 
concentration instabilities responsible for the appearance of waves and oscillations, it is assumed
that the reaction rate $k_3$ depends in a non-linear way on the cytosolic calcium $c$ in the form
\begin{equation}\label{calcio3}
k_3=\kappa_1+\frac{\kappa_2 c^n}{K_d^n +c^n}.
\end{equation}
This simple model provides an excellent quantitative description of experimental cytosolic $Ca^{2+}$ 
oscillations and their periods, and predicts the observed flows in each cycle \cite{friel}. 

\begin{figure}[] % float placement: (h)ere, page (t)op, page (b)ottom, other (p)age
  \centering
  \includegraphics[width=3.5in,height=3.0in,keepaspectratio]{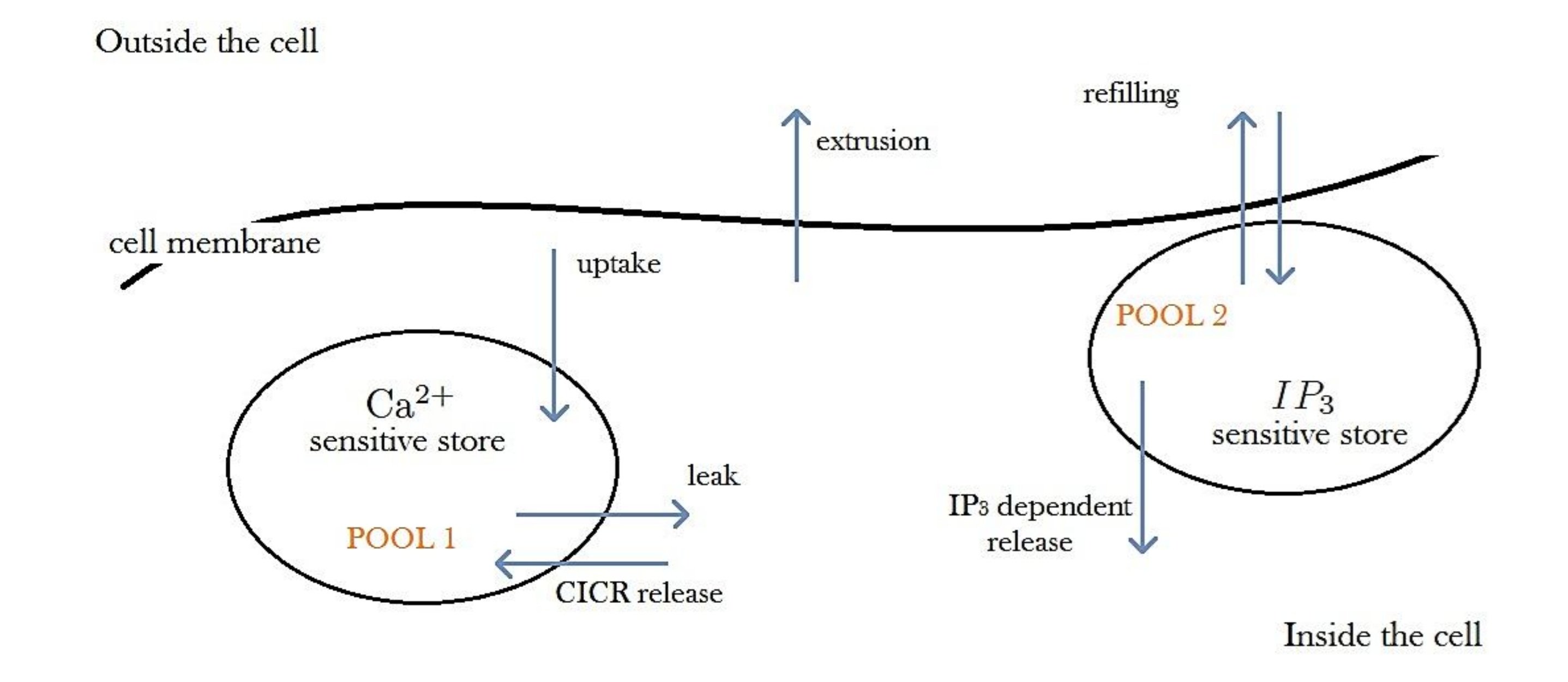}
  \caption{Flux diagram for the two-pool model. In this case, two different calcium reservoirs 
  sensitive to different agonists are activated in series.}
  \label{fig:ondacalciotp}
\end{figure}

%%%%%%%%%%%%%%
\paragraph{The two-pool model.}
%%%%%%%%%%%%%%
The two-pool model, originally proposed by Goldbeter \cite{goldbeter}, suggested the existence of two 
calcium reservoirs that are activated in series by different agonists. Specifically, the first reservoir is 
sensitive to IP$_3$ whereas the second one is sensitive to Ca$^{2+}$. The second pool performs a 
CICR-exchange process with the extracellular medium, see Figure \ref{fig:ondacalciotp}.
Several peculiarities of calcium dynamics for different  cellular types can be successfully reproduced using this model \cite{goldbeter}.

As in the CICR model, a calcium flow balance between pools and the exterior is performed.
However, in this case, the fluxes are given by non-linear relations that are determined independently in the experiment. The important assumption of the model is that the calcium concentration in the
first pool (sensitive to IP$_3$) is constant during the whole process. Thus, as in the CICR model,
only the cytosolic calcium concentration and the concentration inside the second pool $c_s$ 
change with time.

Hence, assuming also that the IP$_3$ gives rise to a constant Ca$^{2+}$ flux $r$, and that the calcium is pumped
to the extracellular medium with rate $-kc$, then the whole dynamics is given by
\begin{equation}\label{eqs:tp1}
\frac{d c}{dt}=r-kc-g(c,c_s),
\end{equation}
\begin{equation}\label{eqs:tp2}
\frac{d c_s}{dt}=g(c,c_s),
\end{equation}

\noindent where $g(c,c_s)$ represents the rate of change of Ca$^{2+}$ in the second pool that depends on 
the uptake and release flows 
\begin{equation}\label{fcc}
g(c,c_s)=J_{\mbox{uptake}}+J_{\mbox{release}}-k_f c_s,
\end{equation}
where
\begin{eqnarray}\label{fcc}
J_{\mbox{uptake}}&=&\frac{V_1 c^l}{K_1^l + c^l}   \\
J_{\mbox{release}}&=&\left( \frac{V_2 c_s^m}{K_2^m + c_s^m}\right)\left(\frac{ c^p}{K_3^p + c^p}\right).
\end{eqnarray}
Here, $J_{\mbox{uptake}}$ is the rate at which the Ca$^{2+}$ is pumped towards the second pool,
that is, by means of an active mechanism, and $J_{\mbox{release}}$ is the rate at which calcium is released from
this second pool. This is an active feedback mechanism for the Ca$^{2+}$ that is essential for the
appearance of calcium oscillations in the cytosol. Finally, $k_f c_s$ is the rate at which the calcium
is released to the extracellular medium. The exponents $l$, $m$ and $p$ and the other parameters 
should be chosen appropriately for each case. 

\paragraph{Calcium waves and the two-pool model.}
Frequently, Ca$^{2+}$ oscillations do not occur in an homogeneous way inside the whole cell. On the contrary,
these oscillations are organized spatially in the form of waves. The velocity of these waves 
is notably similar in many different types of cells (5-20$\mu m/s$). In general, these waves are independent of the
extracellular calcium and they are often concentric, plane or even have a spiral shape. The utility of these
waves can be attributed to the intracellular-extracellular communication.

In order to model these waves, it is assumed that the cytosolic calcium $c$ diffuses, and
therefore it is added a diffusion term to the equation (\ref{eqs:tp1}). Therefore, the resulting 
two-pool model for calcium waves takes the form
\begin{equation}\label{eqs:ondaca1}
\frac{d c}{dt}=D_c\nabla^2 c +r-kc-g(c,c_s), 
\end{equation}
\begin{equation}\label{eqs:ondaca2}
\frac{d c_s}{dt}=g(c,c_s),
\end{equation}
where $D_c$ is the Ca$^{2+}$ diffusion coefficient in the cytoplasm \cite{dupont1994}.

%%%%%%%%%%%%%%%
\subsection{Biochemistry and dynamics of protein motors}
%%%%%%%%%%%%%%%
Kinesins, myosins and dyneis are large families of cytoskeleton-based protein motors that participate 
in the exocytosis-endocytosis cycles of cells, among many other processes, by transporting organelles and vesicles from the inner regions of the cell towards the periphery \cite{vale2003}. 

An important case in which these motors play a key role is in the plasmatic membrane resealing of oocytes.
Here, as we mentioned in the Introduction section, we will consider the model case of membrane
resealing of sea urchin eggs \cite{bi1995, steinhardt1994b}. In this case, it has been suggested that 
protein motors are involved in the docking and delivery of vesicles during exocytosis. 
Using a confocal microscope, in Ref. \cite{gqbi1997} it was observed inhibition of exocytosis in sea urchin 
eggs injected with kinesin and myosin inhibitors. This observations support the hypothesis that kinesin and 
myosin motors mediate two sequential transport steps that recruit vesicles to the release sites of 
Ca$^{2+}$-regulated exocytosis. From this experiment, it was concluded that kinesin leads the slow 
phase of exocytosis delivering vesicles along the microtubules from inner regions of the cell, 
while myosin leads the fast phase, dragging outer vesicles trough filaments of F-actin\cite{gqbi1997}.  It
also exists evidence that slow and prolonged release of insulin in $\beta$-cells is also a process mediated by
protein motors like kinesins and myosins. Finally, it is opportune to mention that theoretical modeling 
offered clear support to the hypothesis that protein motors are responsible for the translocation of 
vesicles containing neurotransmitters in the soma of serotonergic neurons \cite{biophysics}. 

As proteins, enzymes and molecular complexes, the processes developed by protein motors use 
chemical energy stored into molecules such as ATP or GTP, produced by the mitochondrial system 
of the cell \cite{Alberts}. This is an interesting issue: protein and molecular motors
are strongly coupled to their environments, from which they obtain the "fuel" necessary to perform
work during the same process they carry out. That is, molecular machines have not their inner fuel reservoirs.
As a consequence of this, their energetics should be described jointly with that of the surrounding
medium. 

 \emph{In vitro} experimental studies determined the dependence of protein motor activity
on ATP concentrations with well controlled conditions.
This approach has advantages. For instance, it allows for a better analysis of the detailed 
dynamics of the motors irrespective of the medium, but may hide some aspects
of its performance in \emph{in vivo} conditions, such as the transport of proteins, RNA, 
vesicles and even organelles \cite{embjo} that may be related, for instance, to several exocytosis-endocytosis processes \cite{steinhardt1994b,bi1995,biophysics,epithelial,Allosteric,neuro}. 

\paragraph{Protein motor biochemistry.} 
There are essentially two models that represent the motion of protein motors. Here, we will focus our
attention on the so called hand-over-hand model, that was first developed to describe the particular 
problem of intracellular transport via kinesin motion due to ATP 
hydrolysis~\cite{Visscher1999,Visscher2000,ADP,Progresion-Kinesin}.  

The biochemical reaction mechanism modeling the activity of kinesin motors was originally 
proposed in Ref. ~\cite{Visscher2000} and consists on a sequence of
six reactions (see also Refs. \cite{jared2012,jared2014}). The first reaction describes 
the formation of  the enzyme-substrate complex $MKT_\alpha$ due to the capture of an 
ATP molecule ($T$) by the microtubule ($M$) kinesin ($K$) complex $MK$
with the head $\alpha$ of the kinesin attached at position  to the microtubule (the 
head $\beta$ is free in a retrograde position):
\begin{equation}\label{rec:1}\displaystyle%REACCION 1%%%%%%%
MK_\alpha^{}+T\ce{<=>[k_1][k_{-1}]}MKT_\alpha^{}\,.
\end{equation}
In this scheme, the complex $MK$ plays the role of an enzyme that acts over substrate $T$ through a catalytic reaction \cite{Visscher2000}. 

The second reaction is controlled by thermal fluctuations that induce conformational changes of the dimer 
constituting the motor stalk. This produces a secondary enzyme-substrate complex $MKT_\alpha^\prime$
that favors a step forward because corresponds to an advanced position of the $\beta$ head:
\begin{equation}\label{rec:2}\displaystyle%REACCION 2%%%%%%%
MKT_\alpha^{}\ce{<=>[K^{\dag}][]}MKT_\alpha^{\prime}\,.
\end{equation}
In general, this step is influenced by the load $f$ (cargo weight) attached to the kinesin~\cite{Visscher2000}.

The next reaction describes the attachment of the free head domain $\beta$ to the microtubule using an ADP molecule ($D$) and forming the complex $MKT_\alpha D_\beta$. This step is considered the slow step of the reaction mechanism and therefore the one determining the overall 
velocity of the reaction~\cite{Visscher1999}. It is described by
\begin{equation}\label{rec:3}\displaystyle%REACCION 3%%%%%%%
MKT_\alpha^{\prime}\ce{->[k_2][]}MKT_{\alpha}D_{\beta}\,.
\end{equation}
This reaction occurs over a high free-energy barrier that cannot be surmounted by thermal fluctuations.
After this attachment, two reactions occur that prevent the detachment of the 
kinesin from the microtubule
\begin{equation}\label{rec:4}\displaystyle%REACCION 4%%%%%%%
MKT_{\alpha}D_\beta\ce{->[k_3][]}MK_{\beta}T_{\alpha}\,,
\end{equation}
and
\begin{equation}\label{rec:5}\displaystyle%REACCION 5%%%%%%%
MK_{\beta}T_{\alpha}\ce{->[k_4][]}MK_{\beta}(D\circ{P})_\alpha\,.
\end{equation}

The final reaction of the first step corresponds to the hydrolysis of ATP at the active site 
$\alpha$ of kinesin. This reaction produces an ADP molecule and an inorganic 
phosphate $P_i$ \cite{ADP}. Hydrolysis produces a large amount of energy that allows the 
enzyme to liberate the head $\alpha$ and, in this way, surmount the  high free-energy barrier 
associated to step (\ref{rec:3})
\begin{equation}\label{rec:6}\displaystyle%REACCION 6%%%%%%%
MK_{\beta}(D\circ{P})_\alpha\ce{->[k_5][]}MK_{\beta}+D+P_i\,.
\end{equation}
After this reaction, the cycle is completed and one recovers an initial state $MK_\beta$ 
one step forward from the initial position and with lower free-energy
[see Eq. (\ref{rec:1})].  The number of repetitions of this cycle that the molecular motor 
is capable to perform determines its processivity.

\paragraph{ATP dependence of motor velocities.} 
For our purposes, it is now convenient to consider the dependence of protein motors velocity on
the ATP concentration in the surrounding medium. We first recall that  
%For our simplified model, we will suppose that only one motor (representing both kinesin and myosin) is present. We assume that this 
motors are activated when cytosolic calcium in its surroundings reaches some critical value $c_i^c$. 
At this time $t_0$, the motor or the collectivity of motors start their walk along the microtubule 
carrying a cluster of vesicles towards the cell membrane. The time of activation depends upon motor 
position in the cell with respect to the origin of the calcium wave.

Once the motors have been activated, their average velocity determines the rate at which cortical vesicles, 
both docked to the membrane as far from it, reach an fuse with the cell membrane and are exocyted. 
In Ref. \cite{Visscher1999} it was suggested that this dependence of the velocity on ATP 
follows a Michaelis-Menten (MM) kinetics, and therefore the following ATP concentration ([T]) dependence 
\begin{equation}\label{eqs:mm}
v([T])=\frac{v_{max} [T]}{[T]+K_M},
\end{equation}
where $v_{max}$ represents the maximum speed of the motors and $K_M$ is the MM constant of the process. 
We determine both constants  from experiments reported in Ref. \cite{porter1987} for kinesins in sea urchin eggs. 
In these experiments, kinesins adhered to a solid substrate move a filament of microtubule for some 
$\mu$m in presence of several ATP concentrations. The velocity of the microtubule is found to be 
[T]-dependent according to Eq. (\ref{eqs:mm}). In the following section we will use this dependence 
to obtain $v_{max}$ by fitting experimental data with good results.   

\paragraph{ADP inhibition and finite time walks.} 
Several experimental studies showed that ADP may inhibit the processivity of protein motors, 
since kinesin, myosin and dyne in heads act as multi-substrate enzymes 
\cite{ADP, ADP-production-1,ADP-production-2,nature-Inhibition}. 
\begin{figure}[]
\centering
\includegraphics[scale=0.45]{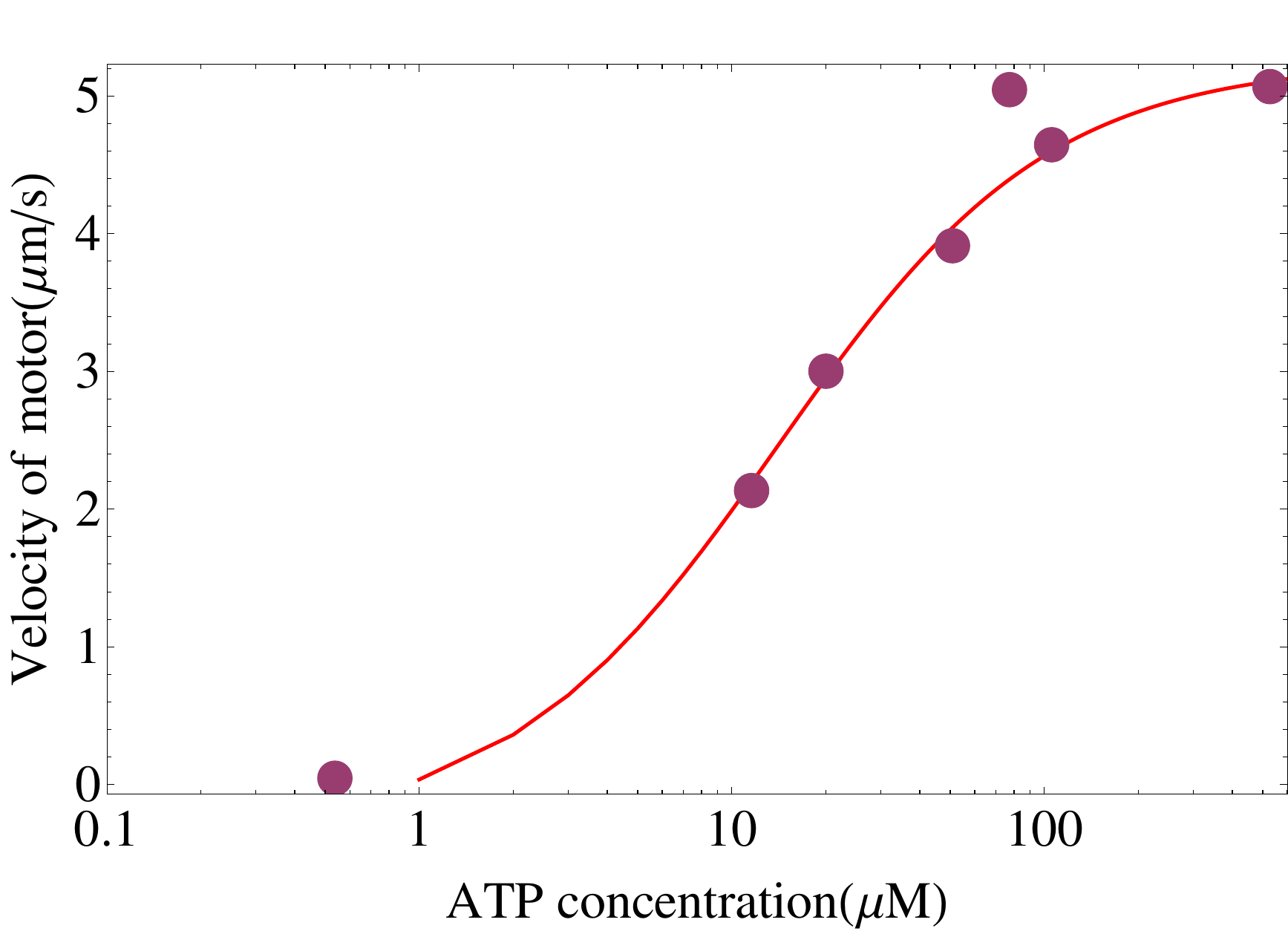} 
\caption{Kinesin velocity as a function of ATP concentration. Fitting was made using Minimal-Squares with a Michaelis-Menten dynamics, eq.(\ref{eqs:mm}). We have found that $v_{max}=5.25 \mu m s^{-1}$ and $K_M=14.93\mu M$.}
\end{figure}

More recently, it was probed theoretically in Refs. \cite{jared2012,jared2014}, that taking into account the 
inhibition by ADP in the reaction scheme already discussed, yields a finite number of steps of a single 
walk of a protein motor. In the case of kinesins, the average number of processive steps predicted was 
around 60 to 120, in excellent accordance with experimental observations.

\begin{figure}[]
\centering
\includegraphics[scale=0.225]{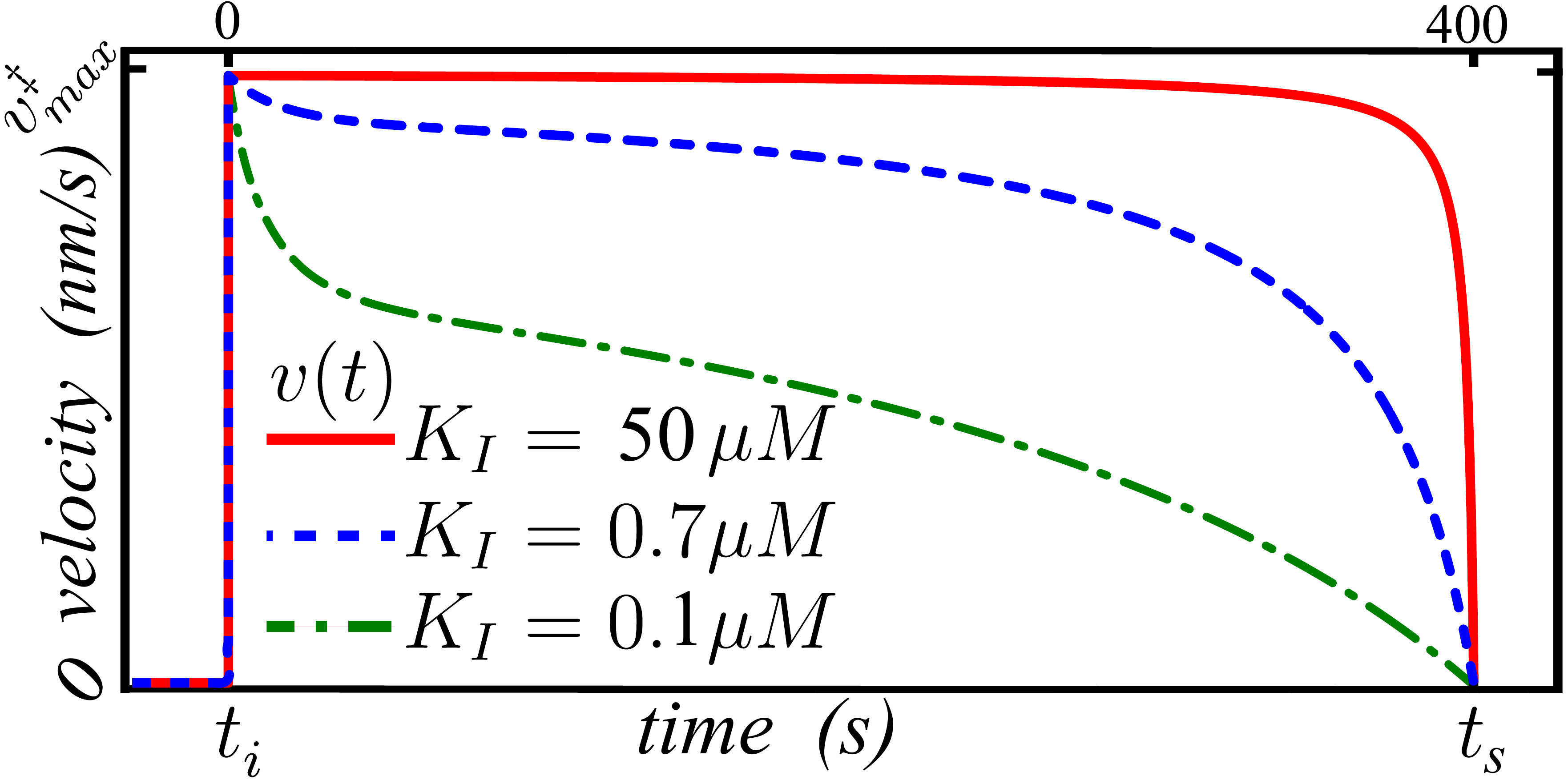}
\caption{Translational velocity of a kinesin motor as a function of time given by Eq. (\ref{eqs:velocity}) with the following values for the constants: values: $\ [T]_o=~1\,mM$,$\ [D]_o=~0$,$\ v^{\ddag}_{max}=900\,nm\,s^{-1}$, $\ k_{cat}=113\,s^{-1}$, $\ k_v=5,600\,s^{-1}$ and 
$\ K_M^{\ddag}=2.24\,\mu{m}$. Except $K^\ddag_M$,  all values were taken from reference~\cite{ADP}. Significant values that do not appear 
explicitly in Eq. (\ref{eqs:velocity}) like $v_{max}=11.25\mu{M}\,s^{-1}$, $[MK]_o=100nM$ 
and $K_{M}^\dag=28\mu{M}$ were taken also from reference~\cite{ADP}.
Figure taken from Ref. \cite{jared2014}.}
\label{Figvelocidadtiempo}
\end{figure}

\paragraph{Time dependence of protein motors velocity.} 
A consequence of this fact is that the velocity of the protein motors is a function of time. This is clear
since each step uses an ATP molecule that, after hydrolysis produces at least one ADP and one $P_i$ 
molecules. This means: 1) The ATP concentration [T] evolves in time following, in first approximation, a 
linear relation \cite{jared2012} 
\begin{equation}\label{T-de-t}\displaystyle
[T]=-v_{max}t - [T]_0, 
\end{equation}
where  [T]$_0$ is the initial ATP concentration. 2) Considering that the autocatalytic production of [ADP] inhibits 
the whole reaction scheme, in particular the formation of the MKT enzyme-substrate complex, 
after many cycles of hydrolysis  the ADP concentration can be modeled by a first order kinetics of the form 
\cite{jared2012,jared2014}
\begin{equation}\label{D-de-t}\displaystyle
[D]_t=[D]_o+[T]_o(1-e^{-k_vt}).
\end{equation}
After solving the corresponding evolution equations of the chemical kinetics, it was shown that 
the velocity of a kinesin activated at time $t_0$ can be written in the form \cite{jared2012}
\begin{equation}\label{eqs:velocity}
v(t)=\frac{v_{max}[\tau- (t-t_0)]}{\alpha-\beta e^{-k_v(t-t_0)}+[\tau- (t-t_0)]},
\end{equation}
where $k_v$ and $\alpha$ are are parameters that can be measured in the experiments
and the stopping time is given by  $\tau=[T]_o/v_{max}$. For the cases when
ATP concentration is low, ADP inhibition controls the dynamics of the kinesin, making its velocity
strongly dependent on time, as it is shown in Figure \ref{Figvelocidadtiempo}. However, for usual
ATP concentrations in living cells, kinesin velocity is practically constant until it stops at a given time
that depends on the initial ATP concentrations and the catalytic constants. In rough terms, the stopping time 
is around $\tau\sim 70 sec$, a time that can be correlated to that for the duration of near pools exocytosis \cite{gqbi1997}. Long time exocytosis may lasts around few hundreds of seconds. 
Here, it should be stressed that the initial time $t_0$ is determined by the initial position of
the molecular motor with respect to position at which the calcium wave initiated, and also by the
calcium wave velocity.
%%%%%

\section{A quantitative example: Early events of fertilization in \emph{sea urchin} eggs}

Fertilization in \emph{sea urchin} eggs has been widely studied by biologists of the last century. There is a great amount of experimental evidence describing multiple processes during activation since the contact of the sperm with the jelly: acrosome reaction, calcium wave spreading over the cell, exocytosis of cortical granules and formation of the fertilization envelope. It is known that a cascade of ionic and metabolic changes occur in the \emph{sea urchin} egg after contact with the sperm \cite{bonder1995}. The egg response to the fertilization stimulus is a rapid depolarization of plasma membrane potential establishing the fast block to polyspermy at the time that the eggs demonstrate a brief global contraction within the first 30 sec after fertilization \cite{eisen1984}. 

Simultaneously, another signaling process induces the production of inositol triphosphate (IP$_3$) and diacylglicerol. An increase in  (IP$_3$) initiates the release of Ca$^{2+}$ from intracellular stores and produces a calcium wave, some seconds after fertilization, that travels across the egg form the site of sperm entry to the opposite pole \cite{stricker1992,whitaker1993}. This wave of elevated calcium propagates trough the entire egg. The maximum of concentration occurs in cortical cytoplasm. This wave is preceded by a distinct, short transient spike of calcium in the cortex which correlates with the time of sperm-egg binding \cite{shen1993}.  

The increase of the intracellular calcium concentration, [Ca$^{2+}]_i$, stimulates in turn a wave of cortical 
granule exocytosis, resulting in the elevation of the fertilization envelope and formation of the hyaline layer 
\cite{eisen1984}. It has been suggested that this transport of vesicles is carried out by protein motors like kinesins and myosins \cite{gqbi1997}. Measurements of secreted hyaline, that is, of fused vesicules in 
the plasma membrane has been reported in \cite{gqbi1997}. %Our purpose here is to give a mathematical model which can give account of those data.

As we have mentioned earlier, kinesin walk is activated and regulated by the presence of ATP 
and intracellular calcium  \cite{gqbi1997}. In fact, after the local increase of [Ca$^{2+}]_i$, kinesins 
and myosins, powered by the hydrolysis of ATP, load the cargo and move along microtubules
%%antes microtubule filaments
  transporting it from the interior of the cell towards the periphery.  Using our multiscale model, 
we show that the calcium wave elevates the [Ca$^{2+}]_i$ allowing activation of kinesins near the 
cortex of the cell and mobilizing cortical granules to the cell membrane where they fuse and spread 
their content to the viteline envelope during the process of exocytosis till fertilization envelope is completed.

%%%%%%CAMBIE EL NOMBRE DE LA SECCION

\subsection{Results}

%%%%%%
\paragraph{Calcium waves.}
%%%%%
During fertilization, after a period of latency which lasts about $\sim$7-40 sec, the eggs of sea urchins generate 
a single $\sim$5-min calcium transient that spreads as a wave across the ooplasm \cite{steinhardt1977}. In \emph{Lytechinus}, induced calcium wave is directly preceded by an influx of calcium leading to a cortical flash  
and is followed by a few post-fertilization calcium transients \cite{shen1993}. This wave spreads throughout 
the entire egg with a spherical wave front and velocity between 5-10$\mu$m/s \cite{stricker1999}. 
\begin{figure}[]
\centering
\includegraphics[scale=0.3]{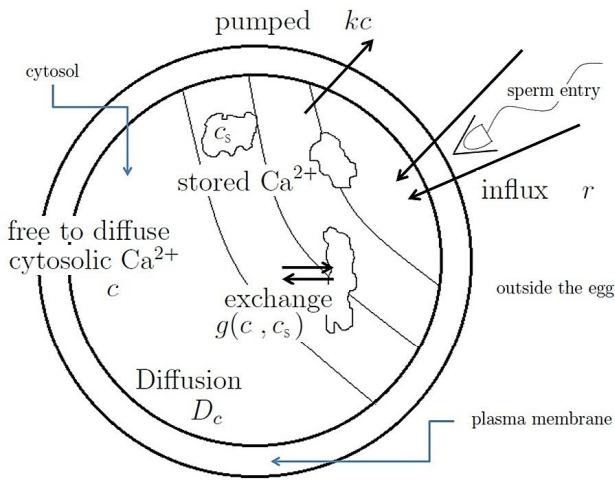}
\caption{Schematic representation of the calcium wave spatio-temporal evolution
according to the two-pool model. Endoplasmatic reticulum is distributed in homogeneous form
helping to maintain and propagate the calcium wave. Cytosolic calcium diffuses with diffusion
coefficient $D_c$.}
\end{figure}\label{fig7}

It is known that external calcium is needed at the onset of sea urchin natural fertilization \cite{Alberts}. This calcium is available from the sea water in the moment of sperm entry during the acrosome reaction \cite{creton1995}. Unfertilized sea urchin eggs can release internal calcium stores via IP$_3$ or non-IP$_3$ -mediated mechanisms \cite{whitaker1993}. Following insemination, a calcium wave occurs in the presence of inhibitors against inositol triphosphate receptors (IP$_3$R s) or ryanodine receptors (RyR's) but not when both inhibitor types are used, suggesting that the two receptors may redundantly participate during fertilization \cite{galione1993}.

Therefore, we will assume that the calcium dynamics within the fertilized egg can be modeled 
with the two-pool extended model. We will suppose that the endoplasmic reticulum (ER) represent the only internal store of calcium  with the release modulated by both IP$_3$ and Ca$^{2+}$. In addition, we will suppose that the network of the ER spreads approximately inside the entire cell \cite{terasaki1991}.
%In addition, we will suppose that the network of interconnected cells of the RE spreads homogeneously inside the cell \cite{terasaki1991}. 
If $c$ and $c_s$ represent concentrations of  cytosolic  and stored $Ca^{2+}$, respectively, and we assume that only the first 
one is free to diffuse at rate $D_{c}$, then the dynamics of calcium is dictated by Eqs.(\ref{eqs:ondaca1}) and (\ref{eqs:ondaca2}). 

\begin{figure}[]
\centering
\includegraphics[scale=.3]{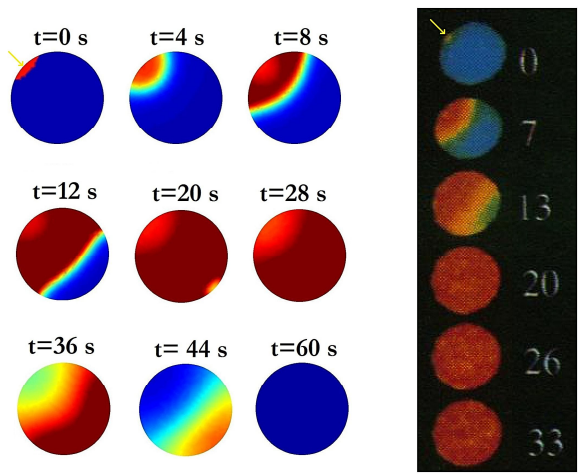}
\caption{The left panel shows the results from our simulation of a calcium wave using the Finite Element method. 
Parameters in equations (\ref{eqs:ondaca1}) and (\ref{eqs:ondaca2}) are given by $r=0.0075\mu$M, $V_1=0.45\mu$ M $s^{-1}$, $k_1=0.9\mu$M, $V_2=2.275\mu$M$s^{-1}$,$k_2=2\mu$M, $k_3=0.9\mu$M, $k=0.04s^{-1}$, $k_f=0.005s^{-1}$,  $D_{c}=50 \mu  m s^{-2}$, $l=2$, $m=2$ and $p=4$. Cytosolic and stored calcium 
are set up at their equilibrium values $c=0.1666$ M and $c_r=2.92$M in almost all the domain, except by a perturbation in the first one of value $1.3\mu$M in the upper left part of the cell. We have used zero flux boundary conditions. Images are taken every 4 sec. Red and blue correspond to maximum and minimum of cytosolic calcium concentration respectively. 
The right panel correspond was taken from Ref. \cite{galione1993} and shows an experimentally observed calcium 
wave within a sea urchin egg.}
\label{fig:calcio2}
\end{figure}

We solved Eqs. (\ref{eqs:ondaca1}) and (\ref{eqs:ondaca2}) using a Finite Element Method. 
As a domain, we used a circle of radius $\sim$50 microns \cite{stricker1999} with a closed border except at the point marked by the arrow in Figure \ref{fig7}, where an initial concentration slightly higher than the equilibrium concentrations simulates the entrance of external calcium due to the
acrosomic reaction. For the equations (\ref{eqs:ondaca1}, \ref{eqs:ondaca2}), we fix typical values 
of the chemical parameters and adjust the coefficients $D_{c}$ and $r$ in such a way 
that: 1) cytosolic calcium spreads as a single wave over the cell, as observed in experiments; 
2) the wave spreads with a circular wave front; and 3) the velocity were 
in accordance with the experiments reported in Ref. \cite{galione1993}, see also Refs. \cite{swan1986, eisen1984, stricker1992, whitaker1993}. We obtained that the two-pool model 
reproduces remarkably well the spatio-temporal evolution of the  
calcium wave in the oopplasm. Our results are represented in the left panel of Figure \ref{fig:calcio2}
where the time evolution of the calcium wave is compared with the results observed in Ref. \cite{galione1993},
right panel. In this figure, at time $t=0$, the small red semicircle indicates the point at which 
takes place the initial local elevation of calcium concentration, that is, where the acrosomic reaction 
takes places. The blue color indicates the low basal calcium concentration inside the oocyte before stimulation.
Front velocity and shape are very similar to the experimental ones. This allows to reproduce the formation
of the vitaline envelope because the activation of the motors occurs at times very similar to the experimental ones. The model does not represents well the time scale for the reduction of calcium concentration which
in the experiments may last few minutes. 

%%%%%%%%
\paragraph{Protein motors activation and dynamics.}
%%%%%%%%
Once having the calcium dynamics, then we proceeded to study the dynamics of protein
motors based on the model discussed in Section II.C.
%{\color{blue} in order of account molecular motor velocity as a function of time}. 
One hypothesis of this model is that the speed has a Michaelis-Menten dynamics type. In the case of kinesin present in the sea urchin egg, this assumption can be corroborated through  {the experiments  reported in Ref. \cite{porter1987}. In the cited work, a partially purified kinesin is adsorbed to a glass coverslip and mixed with microtubules and ATP. The translocating activity of kinesin to microtubules is observed by contrast microscopy at different concentrations of ATP. Experimental data have been fitted with Eq.
(\ref{eqs:mm}) by least squares, corroborating that velocity follows a Michaelis-Menten dynamics. 

This result gives the maximum velocity entering Eq. (\ref{eqs:velocity}) for the speed of kinesin 
as a function of time. In Figure (\ref{fig:perfil}), we show several possible velocity profiles 
varying  two different parameters ($k$, and $v_{max}$). While $v_{max}$  clearly controls the maximum speed reached by a kinesin, $k_v$ indicates how smoothly the kinesin started his walk. In these fittings, 
we have used that average kinesin walk takes about 70 seconds, in accordance with experimental results.  The molecular motor is rapidly activated after calcium increase and starts carrying the transported vesicles 
to the plasmatic membrane of the oocyte. 
\begin{figure}[]
\centering
\includegraphics[scale=0.45]{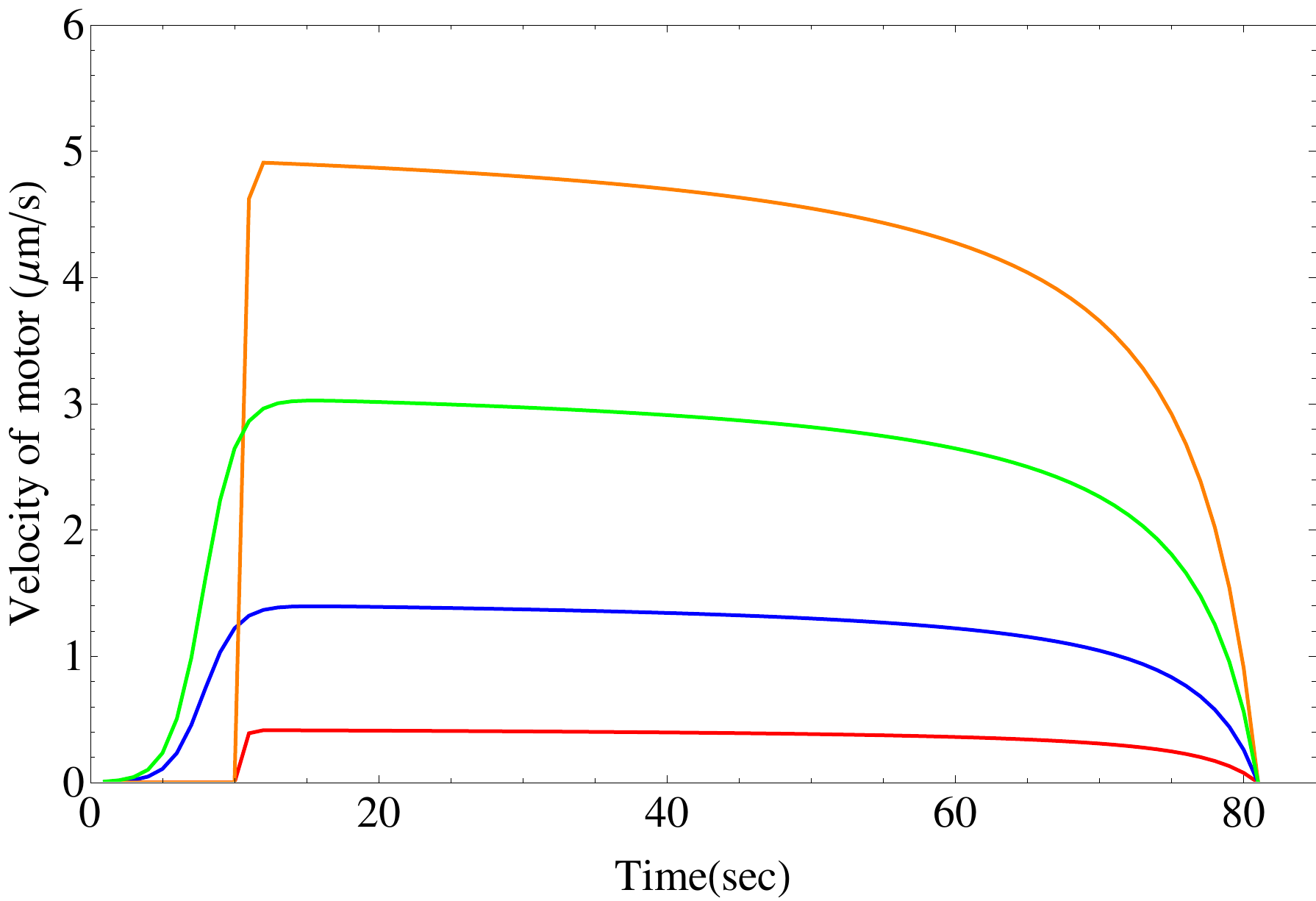} 
\caption{Possible velocity profiles of molecular motor as a function of time. We fix the values of $\alpha=4.8 s$ and $\beta=4.7 s$ in equation (\ref{eqs:velocity}). The values of $k_v$ are $5.6\times10^{3} s^{-1}$ for orange and red lines, and $8.9\times10^{-1} s^{-1}$ for green and blue. Note that a smaller value of this parameter softens engine acceleration at the beginning of its walk. The values of $v_{max}$ are 5.25, 3.25, 1.5 and 0.44 micrometers per second from top to bottom.}
\label{fig:perfil}
\end{figure}

It is convenient to stress that the time at which a motor starts its walk depends on 
its distance from the point of fertilization and to the cell cortex. Vesicles near the entry point of the sperm come quickly to the plasmatic membrane. Those initially located far away from the
stimulation point have a larger latency for the initialization time $t_i$ that depends on the calcium wave 
velocity and shape. The important result represented in Figure (\ref{fig:perfil}) is that, given the ATP
ooplasm conditions, the velocity of the protein motors is almost constant within the duration of the 
exocytosis. Note from Figure \ref{Figvelocidadtiempo}, that for large ADP concentrations present, 
the velocity profile as a complicated time dependence, see for instance, the dashed-dotted line.

%%%%%%%
\paragraph{Directed diffusion and cumulative exocytosis.}
%%%%%%%

\begin{table}
\begin{center}
    \begin{tabular}{| c | c | c | c | c | c | c |}
    \hline
    color  & $D_v (\mu m^2 s^{-1})$ & $v(\mu m s^{-1})$ & $d (\mu m)$ & $n$ \\ \hline
   green  & 0.6  & 1.0 &  5  & 47  \\ \hline
  orange  & 0.7  & 0.2  &  8 & 76 \\ \hline
    blue  & 0.6  & 0.05  & 8 & 27  \\
    \hline
    \end{tabular}
\end{center}  
\caption{Parameters of the model for the three profiles of Figure \ref{cumulative}. See text for details.}
\end{table}

Finally, we quantify the results of cumulative exocytosis reported in \cite{gqbi1997}
by linking the previous theoretical results through the diffusion model. In these experiments, the number of vesicles that are secreted is counted as a function of time over the 
plasmatic membrane of the egg using confocal microscopy.  Experiments show that decreasing 
the external calcium causes a slower rate of exocytosis. These results can be understood, at least qualitatively, using the diffusion model discussed in Section \ref{sec:difusion}. 

In view of the results already discussed for the time dependence of motor velocity, we may 
simplify the numerical resolution of the diffusion equation by assuming, in first approximation, 
that kinesin moves with a constant velocity. Then we solve numerically equation (\ref{eqs:fp}) and
reproduced the experimental data on cumulative exocytosis by adjusting the parameters (diffusion coefficient, 
friction coefficient, frequency, initial position of the vesicles and motor velocity). 
As it is shown by the lines in Figure \ref{cumulative}, the fits of the cumulative exocytosis with this model is excellent, 
and gives many important information on vesicle distribution, motor speeds and the viscoelastic properties of the cell that otherwise remain hidden. This information is contained in Table I, where the values of the diffusion coefficient,
the average protein motor velocities, the initial distance between the clusters of vesicles with respect to the
plasmatic membrane and the number of vesicles per cluster are reported. These values correlate well with those obtained
in other works (\cite{biophysics}), although in the present case the average velocity is much higher, fact possibly
related with the biological nature of the response. The number of vesicles per cluster is only indicative and 
approximate, since no direct comparison with the ultrastructure of the cell allows to compare these values. In addition,
fluorescence measurements should be normalized to a given region and the efficiency of vesicle fusion with the
membrane was assumed equal to one.

\subsection{Discussion}

The fits of the calcium wave using the two-pool model (with the inclusion of the diffusive term) correctly reproduce the profile and speed of the calcium wave in its initial propagation through oocyte. Notwithstanding, for the recovery period in which calcium concentration decreases to its basal 
value we have found (of about 60 seconds) is well below that found in fluorescence experiments of about 7 minutes \cite{stricker1992}.  This can be improved in our model considering the continuous exchange of calcium with the outside and not only at the time of the initial application. Furthermore, in our present work, the stores of calcium inside the cell are distributed homogeneously within the cell, and therefore could no replicate the rapid increase of calcium near the cell membrane at the moment of fertilization \cite{shen1993}. We thought this can be due to the greater presence of calcium stored in outer regions of the egg. However, for qualitative purposes, the occurrence of calcium waves founded with the two-pool model helps to understand the exchange mechanism which produces calcium waves inside the egg. 

We have satisfactory found that engine speed obeys a Michaelis-Menten dynamic, corroborating that a kinesin can reach a speed of up 5.25$\mu m s^{-1}$ in \emph{in vitro} experiments with high concentrations of ATP. However, the models presented in this work allow us to think that this velocity may be much lower in the fertilized sea urchin eggs, where the engine speed when carrying vesicles is severely limited not only 
by the availability of intracellular ATP \cite{jared2014}, but also by the conditions of friction or drag of the intracellular environment where it moves \cite{santamaria2009,biophysics}. It is therefore not surprising that in our {fittings} of cumulative exocytosis, the effective velocity  we found on our adjustments (between 1 and 0.05$\mu m s^{-1}$) is substantially lower than that of \emph{in vitro} experiments . For the purposes of this model, we assumed a constant motor speed, however, more accurate results may be obtained following
the time dependence reported in \cite{jared2014}. This may allow one to understand that this is only an approximation for intracellular processes with a time scale considered comparably high to the acceleration and deceleration of kinesin. However, these speed changes can be crucial in terms of energy to understand the processivity and stop of these motor proteins. 

The diffusion model and drag of the transported vesicles within a cell proposed by Santamaría-Hólek in \cite{biophysics} proves to be extremely successful in quantifying cumulative exocytosis in eggs of sea urchins. However, when coupled with the calcium wave patterns, processivity of molecular motors, applicability of this model to this biological system could be expanded to explain and quantify the formation vitaline envelope emerging into the egg to prevent polyspermy. This is due to the fact that the model perfectly differentiate the start and coupling of each process: when the fertilization process starts, it emerges a wave of calcium whose concentration is known at each position as a function of time \cite{dupont1994}; when the calcium concentration reaches locally a critical value at time $t_i$, a molecular motor begins its walk dragging outward a vesicle \cite{jared2014}; the rate and the number of vesicles arriving at each position can be determined by the conditions of diffusion and drag of the medium \cite{biophysics}. Thus, excepting the numerical complexity of coupling all the required equations, the above model would allow us in the future to quantify the spatio-temporal distribution of secreted vesicles and understand the formation of the envelope.

%%%%%%%
\section{Conclusions.}

The multi-scale models are very useful when trying to describe processes whose causes are described in different spatio-temporal scales, fact which occurs in most biological processes. In this paper, 
we have revisited the description and modeling of three particular physicochemical mechanisms 
that are essential in secretion processes: a) formation of calcium waves, b) the dynamics of molecular motors based on its biochemistry, and 3) directed diffusion of vesicles within a fertilized egg. 
%This aim to integrate these models and show their connection in a complex process such as exocytosis. 

Within this analysis and with the help of previously published results in literature, it was found that the extended two-pool model can be used to understand the propagation of calcium waves in the ooplasm of  
sea urchin eggs. Furthermore, the approach we followed in our work has the advantage of allowing the integration of specific information of other cell types where these waves occur, for instance: geometry, exchange flows with the extracellular medium, location of internal pools and intensity of initial application. We also reviewed and used  a model for the processivity of protein 
motors which is based on their biochemistry. This model yields the time dependent velocity profiles 
for the kinesin, protein which plays a key role in the translocation of clusters of vesicles in many
cell types. For this, we reviewed and solved a simple diffusion model for vesicle transport that integrates various factors affecting the movement of vesicles before fusing with the cell membrane, namely diffusion and drag forces due to the activity of protein motors. By solving this three
component model, we have successfully quantified cumulative exocytosis previously reported in literature. 

More importantly, in this paper we have described how these three schemes can be integrated to quantify with accuracy the formation of the viteline envelope. For this, in addition to the description of each model, we proposed the relationships between each of them, relationships which are essential to understand this complex process. Thus, this work establishes a theoretical foundation useful for the understanding and 
analysis of secretion processes, whose difficulty now is merely based on the numerical resolution of mathematical equations.

In future work, we pretend to propose a more detailed quantitative model which will consider aspects related to the kinetics of docking and priming of the vesicles in the porosomes \cite{anderson,jena}, as well as the detailed mechanical description of swelling of the vesicles and the expulsion of its content.

\section{Acknowledgements}
We acknowledge N. J. L\'opez-Alamilla for useful discussions. ALD acknowledges CONACyT
for financial support under fellowship 221505 and DGAPA UNAM through the grant No. IN-113415.

% Termina el documento
\end{document}